\documentclass[twocolumn]{aastex701}
\shorttitle{X-ray activity in TYC 4144-329-2}
\shortauthors{G\"unther et al.}
\graphicspath{{./}{figures/}}

\begin{document}

\title{Stellar activity in a post-merger Giant star}

\author[orcid=0000-0003-4243-2840,gname='Hans Moritz',sname='Günther']{H. M. G\"unther}
\affiliation{MIT Kavli Institute for Astrophysics and Space Research, 77 Massachusetts Avenue, Cambridge, MA 02139, USA}
\email[show]{hgunther@mit.edu}  

\author[orcid=0000-0001-9834-7579,gname='Carl',sname='Melis']{C. Melis}
\affiliation{Department of Astronomy \& Astrophysics, University of California San Diego, La Jolla, CA 92093-0424, USA}
\email{cmelis@ucsd.edu} 

\author[orcid=0000-0002-9632-6106,gname='Ken J.',sname='Shen']{K. J. Shen}
\affiliation{Department of Astronomy and Theoretical Astrophysics Center, University of California, Berkeley, CA 94720, USA}
\email{kenshen@astro.berkeley.edu} 

\author[orcid=0000-0002-9405-8435,gname='Eric G.',sname='Blackman']{Eric G. Blackman}
\affiliation{Department of Physics and Astronomy, University of Rochester, Rochester, NY 14627, USA}
\email{blackman@pas.rochester.edu}

\author[orcid=0000-0002-8636-3309,gname=Keri,sname=Hoadley]{K. Hoadley}
\affiliation{University of Florida, Department of Astronomy, Bryant Space Science Center, Gainesville, FL, 32611 USA}
\email{khoadley@ufl.edu}

\author[orcid=0000-0002-5094-2245,gname='P. Christian',sname='Schneider']{P. C. Schneider}
\affiliation{Christian-Albrechts-Universität zu Kiel, Institut für Theoretische Physik und Astrophysik, Leibnizstr. 15, 24118 Kiel, Germany}
\email{astro@pcschneider.eu}

\begin{abstract}

Binary stars are the progenitors of exotic objects, such as supernovae and gravitational wave sources. 
However, some systems merge earlier when the primary expands into a giant star at the end of its lifetime on the main sequence
and causes its companion to spiral in.
TYC~4144-329-2 is a candidate post-merger first ascent giant star with a circumstellar disk seen in the infrared (IR).
We detect weak X-ray and far ultraviolet (FUV) emission from TYC~4144-329-2, which we interpret as a signature of weak coronal activity.
We also find variability in H$\alpha$ line profiles and in the optical light curves and we suggest an intermediate geometry where 
the line-of-sight passes through the upper layers of a flared disk with a time variable column density.
While the X-rays must be coronal, the H$\alpha$ line profiles point to ongoing and variable accretion.
We suggest that TYC~4144-329-2's merger was more recent than other stars in this class and that it did not yet have time to develop a deep convection zone.
If this scenario holds, TYC~4144-329-2 would be a unique probe of the earliest stages in the evolution of post-merger systems.

\end{abstract}


\section{Introduction}
\label{sect:introduction}

One third of all stars are in binary or multiple systems \citep{Raghavan+2010}. 
If the components in such a system are close enough to interact, they can go through very different evolutionary pathways than single stars do. 
Many aspects of interacting binaries are subject to intense study, such as LIGO detections of neutron star or black hole mergers. However, stellar mergers can also happen in non-degenerate binary stars if they are in a close orbit. Given the high incidence of binaries, understanding stellar binary mergers is important for a large number of stars and also limits the number of systems available for interaction in later stages (neutron stars, black holes).
While many details remain unclear, recent simulations and observations suggest the following scenario \citep{2020Natur.587..387H}: The primary evolves off the main-sequence (MS) first because it is more massive. As it ascends the giant branch, it expands, overflows its Roche lobe and forms a disk around the binary. At the same time the secondary loses orbital angular momentum until it merges with the primary. The system launches outflows from the disk which are visible for a few thousand years. The merger remnant has a larger radius, luminosity, and additional angular momentum compared to a giant that did not accrete a companion, all of which leads to a stronger convection zone and a more active dynamo compared to a single star without a merger history, generating magnetic fields which power a chromosphere and heat a corona to the point where it emits X-rays. As the star contracts, it spins up and its ratio of X-ray to bolometric luminosity $L_\mathrm{X}/L_\mathrm{bol}$ increases while the outflow fades. In effect, the additional angular momentum accreted during the merger rejuvenates the solar-like activity as in \object{FK Com}, a giant star with a convective envelope and a strong dynamo \citep{2016ApJS..223....5A}. There is no disk around FK~Com today, but its high X-ray activity can be explained if it was rejuvenated in a merger event in the past.

This idea is based on observations of a small group of suspected recent stellar merger remnants known today. Many of those stars are hidden behind significant gas and dust column densities \citep{2020RNAAS...4..238M}, but examples of stars that have been studied are TYC 4144-329-2 \citep{2009ApJ...696.1964M}, \object{HD 233517} \citep{2003ApJ...582.1032J}, and \object{TYC 2597-735-1} \citep{2020Natur.587..387H}. 
\object{BP Psc} is often included in this group \citep{Zuckerman_2008,2010ApJ...719L..65K} but recent proper motion studies of its jet identify it as a young, close-by object instead of a farther-way post-merger giant \citep{2022MNRAS.516.5863P}.
In the scenario described above, TYC~2597-735-1 would represent a stage just a few thousand years after it destroyed its close companion, since it shows a circumstellar disk observed as an infrared (IR) excess over the photospheric emission and outflows have been resolved in imaging. TYC~2597-735-1 is detected in X-rays \citep{2022AJ....163..173G}. 

There is actually a class of objects where the stellar merger is directly observed. In our Milky Way that includes \object{V383 Mon} \citep{2003ApJ...582L.105S,2005A&A...436.1009T}, \object{V4332 Sag} \citep{1994IAUC.5942....1H}, and in particular \object{V1309 Sco} \citep{2008IAUC.8972....1N,2011A&A...528A.114T,2019MNRAS.486.1220F}. The latter had an outburst of $>7$~mag in 2008. Retrospective analysis of the pre-merger light curve identified it as a contact binary with an orbital period of $\approx1.4$~d, decreasing with time. It is not clear how this relates to objects like TYC~2597-735-1 that we observe thousands of years after their merger; those might be older objects of the same class or follow a different merger pathway.

In this work, we study \object{TYC 4144-329-2}, a suspected post-merger first-ascent giant branch star that is surrounded by a dust and gas disk. The disk is bright in the $K$ band and at longer wavelengths \citep{2009ApJ...696.1964M}. Its spectral energy distribution (SED) mimics the SEDs of young stars with proto-planetary disks. 
The disk around TYC~4144-329-2 has a luminosity close to one fifth of the stellar luminosity. A flat disk does not intercept enough stellar light to re-radiate that much emission, so the disk must be flared.
Black-body fits to the IR excess indicate a disk radius around 30~AU. The stellar spectrum of TYC~4144-329-2 shows emission lines which are associated with accretion in young stars, so it is likely that the inner edge of the disk is connected to the star and material falls in \citep{2009ApJ...696.1964M}. 

TYC 4144-329-2's binary companion TYC 4144-329-1 is a subgiant and indicates an age far beyond the lifetime of proto-planetary disks \citep[$<10$~Myr, ][]{2026A&A...707A.216P}. The most plausible pathway to form a disk around an evolved star such as TYC~4144-329-2 is the consumption of a short-period companion, either a hot Jupiter or a low-mass stellar companion. Depending on the orbital parameters and evolutionary pathway, the companion might have been destroyed in the process \citep{2003ApJ...582.1032J} or could just be stripped of its outer atmosphere with a core that continues to orbit within the current gas and dust disk \citep{2003ASPC..293...76W}.
The accretion of a companion changes the properties of the atmosphere of the primary star. A lack of Li in TYC 4144-329-2 \citep{2009ApJ...696.1964M} indicates that the accreted object may itself not have been Li-rich, i.e.\ either it was massive enough to burn through its own Li (more massive than a brown dwarf) or it simply did not have enough mass to significantly pollute the phototosphere of TYC 4144-329-2 (e.g., a planet). The stellar merger process can lead to the formation of convective zones in the photosphere \citep{2007MNRAS.375..909S} and a strong magnetic field \citep{Schneider+2016}, which power a convective dynamo and thus produce photospheric spots.

\cite{2009ApJ...696.1964M} characterized the system spectroscopically and derived parameters for both stars. Since that work, however, the Gaia mission has provided an accurate parallax to both stars at a considerably closer distance. In section~\ref{sect:thetyc41443291andtyc41443292binarysystem} we rederive system parameters based on the new parallax and literature values for other photometric and spectral properties of TYC~4144-329-1 and TYC~4144-329-2. We proceed to describe new observations of X-ray, UV, and optical data in section~\ref{sect:observationsanddataanalysis} before we present our results in section~\ref{sect:results}. We discuss the activity of TYC 4144-329-2 in section~\ref{sect:discussion} and end with a summary in section~\ref{sect:summary}.

\section{The TYC 4144-329-1 and TYC 4144-329-2 binary system}
\label{sect:thetyc41443291andtyc41443292binarysystem}

TYC~4144-329-1 and TYC~4144-329-2 have Gaia distances compatible within the uncertainties with an average of $385\pm2$~pc \citep{2016A&A...595A...1G,2023A&A...674A...1G}, which we adopt as the distance to the system. This distance is considerably closer than the value of 550~pc estimated by \citet{2009ApJ...696.1964M} and thus we rederive stellar and system parameters here. 
This system is a wide (5.5\arcsec, about 2000~au) binary but may have been a hierarchical multiple in the past before the primary star TYC~4144-329-2 accreted a short-period companion. 

We set the following priors for the fit: Effective temperature $T_\mathrm{eff}=4800\pm100$~K and metallicity $[Fe/H]=-0.2$ for TYC 4144-329-1 from \citet{2009ApJ...696.1964M} and $B, V$ from \citet{2000A&A...355L..27H}, $G$ from Gaia, and $J,H,K$ from 2MASS \citep{2006AJ....131.1163S}. We fit MIST evolutionary models \citep{2016ApJ...823..102C,2016ApJS..222....8D} with the \texttt{isochrones} package \citep{2015ascl.soft03010M}. The fit is fully bayesian \citep{Buchner+2014}, but all relevant posteriors turn out to be single-peaked and almost symmetric, thus we report the mean and standard deviation as values in Table~\ref{tab:stellar_parameters}. We then take the best-fit age for TYC 4144-329-1 as the age of the system and use it as prior to fit TYC 4144-329-2. 
\citet{2009ApJ...696.1964M} show that the SED of TYC 4144-329-2 is already dominated by the disk in $H$ band, so we only use $B, V, G, J$ magnitudes and increase the uncertainties to 0.2~mag to account for variability (see discussion in section~\ref{sect:opticalvariability}). 

\begin{table*}
\caption{Stellar parameters of TYC 4144-329-1 and TYC 4144-329-2.\label{tab:stellar_parameters}}
\centering
\begin{tabular}{lccr}
\hline
Parameter & TYC 4144-329-1 & TYC 4144-329-2 & reference \\
\hline\hline
\multicolumn{4}{c}{Input parameters for the isochrone fitting} \\
\hline
$T_\mathrm{eff}$ [K] & $4800 \pm 100$ & $7000 \pm 200$ & (1)\\
$[Fe/H]$ & $-0.2 \pm 0.1$ & $-0.2 \pm 0.1$         & (1)\\
parallax [mas] & \multicolumn{2}{c}{$2.56 \pm 0.02$} & (2)\\
$B$ [mag] & $11.01 \pm 0.04 $ & $11.4 \pm 0.2 $   & (3)\\
$V$ [mag] & $10.11 \pm 0.02 $ & $10.7 \pm 0.2 $   & (3)\\
$G$ [mag] & $9.8622 \pm 0.003 $ & $11.0 \pm 0.2 $ & (2)\\
$J$ [mag] & $8.47 \pm 0.03 $ & $9.2 \pm 0.2 $     & (4)\\
$H$ [mag] & $8.04 \pm 0.02 $ & -- & (4)\\
$K$ [mag] & $7.94 \pm 0.03 $ & -- & (4)\\
age [Gyr] & -- & $2.4 \pm 0.6 $ & (5) \\
\hline
\multicolumn{4}{c}{Derived parameters from the isochrone fitting} \\
\hline
$T_\mathrm{eff}$ [K] & $5000 \pm 30$ & $7000 \pm 200$ & \\
age [Gyr] & $2.4 \pm 0.6$ & $1.5 \pm 0.5$ & \\
$M_\star$ [$M_\odot$] & $1.5 \pm 0.1 $ & $1.7 \pm 0.1$ & \\
$R_\star$ [$R_\odot$] & $4.95 \pm 0.05$ & $2.6 \pm 0.3$ & \\
$L_\star$ [$L_\odot$] & $13.8 \pm 0.3$ & $15 \pm 4$ \\
$\log g$ [cgs] & $3.22 \pm 0.04$ & $3.8 \pm 0.1$ & \\
$A_\mathrm{V}$ [mag] & $0.04 \pm 0.03$ & $1.1 \pm 0.3$ & \\
\hline
\end{tabular}
\tablecomments{
(1) \citet{2009ApJ...696.1964M}
(2) \citet{2023A&A...674A...1G}
(3) \citet{2000A&A...355L..27H}
(4) \citet{2006AJ....131.1163S}
(5) from fit of TYC 4144-329-1}
\end{table*}
Given the smaller distance, it is not surprising that our fits result in less luminous and lower mass stars than derived by \citet{2009ApJ...696.1964M}. 
As a consistency check, we can look at the derived values for $\log g$, which we obtain as output and compare them to the values determined from spectral lines by \citet{2009ApJ...696.1964M}. For TYC 4144-329-1, we find $\log g = 3.22\pm0.04$ consistent with the luminosity-sensitive lines that place the star between a dwarf and a giant. While the fit result for the age of TYC 4144-329-2 comes out slightly lower than the input prior, it is still compatible with the physically sensible scenario of a co-eval binary. Since the stellar merger pathway might change the properties of TYC~4144-329-2 compared to a single star, we continue assuming the age of TYC~4144-329-1 as the system age. On the other hand the current effective temperature, mass, radius, and luminosity depend directly on the observed magnitudes and are robust even if the evolutionary models do not account for the merger history.
TYC~4144-329-2 is also clearly evolved off the main-sequence. There is no constraint on the surface gravity from the spectra because of the larger rotational velocity \citep{2009ApJ...696.1964M} and $A_\mathrm{V}$ might actually be time variable (see section~\ref{sect:opticalvariability}) influencing our fit done on non-simultaneous photometry.

\section{Observations and data analysis}
\label{sect:observationsanddataanalysis}

We begin by describing the high-energy photometry from Chandra (section~\ref{sect:chandra}) and GALEX (section~\ref{sect:uvdata}). Then we turn to optical data starting with KWS ground-based lightcurves in section~\ref{sect:kwslightcurves} followed by TESS lightcurves in section~\ref{sect:tesslightcurves} and DASCH lightcurves in section~\ref{sect:otherlightcurves}. Finally, we describe our new ground-based spectroscopy in section~\ref{sect:opticalspectroscopy}.

\subsection{Chandra}
\label{sect:chandra}

This research employs a list of Chandra datasets, obtained by the Chandra X-ray Observatory, contained in~\dataset[DOI: 10.25574/cdc.678]{https://doi.org/10.25574/cdc.678}.
TYC 4144-329-1 and TYC 4144-329-2 were observed using the ACIS-S detector for 30.6~ks split over OBSID 26495 on 2024-04-17 for 17.4~ks and OBSID 29375 on 2024-04-18 for 13.2~ks.  The data were processed using CIAO 4.17 \citep{2006SPIE.6270E..1VF,2026arXiv260514144F}.


Visual inspection of both event files shows a few sources in the field with coordinates that match between observations. All those sources have few counts, so formal fitting of coordinate positions does not refine the relative astrometry. Only a single source can be readily identified with a known catalog source (\object{FIRST J102225.4+613031}) but is far off-axis. Its coordinates match the catalog position within the uncertainty but are not good enough to be used for astrometric correction due to Chandra's PSF which rapidly degrades off-axis. 

\begin{figure}
\plotone{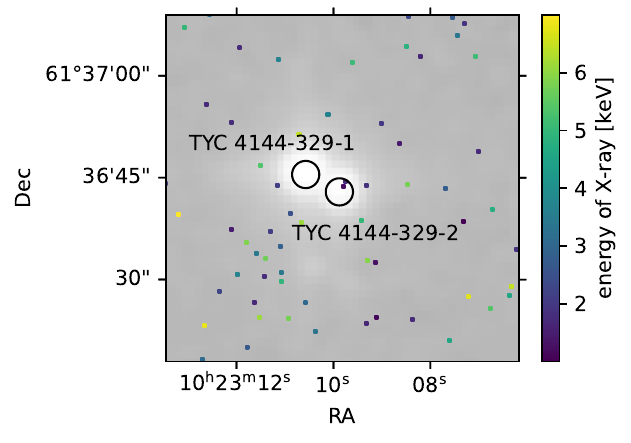}
\caption{Small squares show the positions of X-ray photons detected by Chandra. They are color-coded according to the event energy. Black circles with a radius of 2\arcsec{} mark the source extraction regions for the two stars, centered on their optical location at the time of the Chandra observation. The background image shows the 2MASS $J$ band image of the field \citep{2006AJ....131.1163S} from \dataset[10.26131/IRSA121]{http://dx.doi.org/10.26131/IRSA121}.\label{fig:chandra_photons}}
\end{figure}
We use a large, apparently source-free region on the same chip to define our background.
Over the entire merged observation in the energy range 1.0-5.0 keV we find a background of 0.014 counts per pixel.

The number of counts required to claim a source detection in a given region depends on two parameters: The maximum probability we allow for a false detection (false positive) that can happen when a background fluctuation produces enough counts to cross our detection threshold, and the probability that a source of a given flux will be missed because the Poisson randomness of the source flux happens to produce a count number below the detection threshold (false negative). We set the probability of a false positive to 95\% and the probability of a false negative to 50\%. Following the Baysian algorithm of \citet{2010ApJ...719..900K} as implemented in the CIAO tool \texttt{aplimits}, we find that the minimal number of counts required to claim a detection in a circle with radius 2\arcsec{} is 1 count. Of course, over the entire field-of-view this approach would yield some spurious detections if applied blindly, but for a given source location (in our case the known locations of TYC~4144-329-1 and TYC~4144-329-2) a single count indicates a detection with 95\% significance - a testament to Chandra's sensitivity and low background rate. An undetected source has an upper limit of 1.45~counts, i.e.\ a source with a higher flux is expected to produce at least one count given our choice for the false negative limit. 

Figure~\ref{fig:chandra_photons} shows the positions of the detected X-ray photons in the field. TYC~4144-329-1 is undetected with an upper limit of $<5\;\times 10^{-5}$ counts~s$^{-1}$, and TYC~4144-329-2 is detected with 2 counts in our 30.6~ks exposure. Obviously, the relative uncertainty on the flux of TYC~4144-329-2 is large; we evaluate confidence intervals following \citet{2014ApJ...796...24P} as implemented in the CIAO tool \texttt{aprates}.

Accounting for the background, the best estimate for the count rate of TYC~4144-329-2 is 1.9 counts in 30.6~ks, with a lower limit of 0.8 and an upper limit of 3.8 counts (68\% credible interval).

The energies of the two detected photons are approximately 1.2 and 1.7~keV.

\subsection{UV data}
\label{sect:uvdata}

GALEX \citep{2005ApJ...619L...1M} observed the field in the FUV band (1350-1750~\AA{}) on 2004-01-21 for 199~s and in the NUV band (1750-2800~\AA{}) for 119~s. A second, longer NUV observation was done on 2005-01-14 for 1700~s. We retrieve the photon list for these observations using the \texttt{gPhoton} package \citep{2016ApJ...833..292M} \dataset[10.17909/T9CC7G]{http://dx.doi.org/10.17909/T9CC7G} and combine the two NUV observations.

\begin{figure}
\plotone{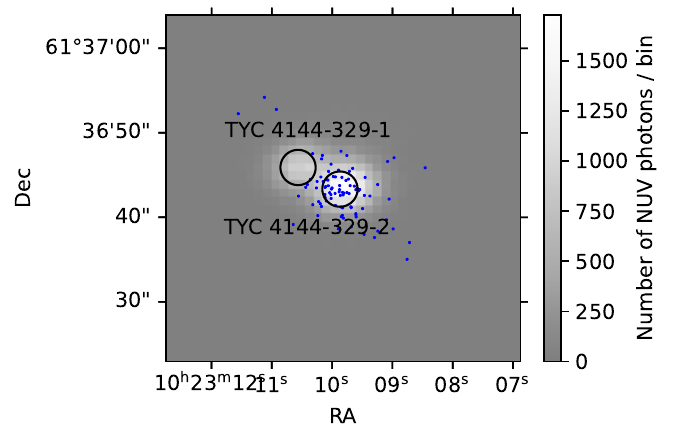}
\caption{{Background image: Binned NUV photon counts from GALEX.
Blue dots mark individual FUV photons. There are fewer FUV than NUV photons because the FUV exposure time is shorter, but the sources are also fainter in the FUV.
Both sources are detected and can be separated in the NUV.
In the FUV, photons cluster around TYC 4144-329-2.
The black circles mark the $1\sigma$ radius (2.1\arcsec) of a Gaussian that represents the NUV PSF.}
\label{fig:galex_photons}}
\end{figure}
Figure~\ref{fig:galex_photons} shows NUV and FUV photons detected by GALEX. In the NUV both TYC~4144-329-1 and TYC~4144-329-2 are detected; in the FUV only TYC~4144-329-2 is seen. We do not observe any significant variability within the second NUV observation (the observations in 2004 are too short to test for variability). The NUV spatial distribution is compatible with GALEX point sources; the radial distribution of the FUV emission shows a small bump that would also be consistent with an extended or a weak additional source about 6-8\arcsec{} to the south-west of TYC~4144-329-2. A KS-test gives a probability of just 2\% that the observed FUV distribution is consistent with the GALEX point-spread-function (PSF) for the FUV. However, TYC~4144-329-2 is located close to the edge of the FUV detector, while the PSF given by \citet{galexdoc} is axisymmetric because it is averaged over many sources. This is problematic because the GALEX PSF is generally slightly elliptical. Thus, we regard this result as tentative, despite the formally convincing KS-test result.

To avoid source confusion, we extract aperture photometry in a small circle of 1.5\arcsec{} for the NUV, but use a larger area for TYC~4144-329-2 in the FUV marked by the black circle in Figure~\ref{fig:galex_photons}. The background is determined from an annulus with inner radius 20\arcsec{} and outer radius 30\arcsec{} centered on the source. We correct the magnitudes found in aperture photometry for the aperture size using the tables of \citet{2007ApJS..173..682M}. Since TYC~4144-329-2 is much brighter than TYC~4144-329-1 in the FUV, the upper limit on the FUV flux is determined mostly by the flux from TYC~4144-329-2 at the location of TYC~4144-329-1 which acts as a background. We use the black circle in figure~\ref{fig:galex_photons} as source region and an annulus around TYC~4144-329-2 with the inner and outer radius matching the closest and farthest distance of the black circle from TYC~4144-329-2 as background region; we exclude a part of the annulus close to TYC~4144-329-1. We then use the algorithm from \citet{2014ApJ...796...24P} again to determine confidence bounds. The resulting numbers are listed in table~\ref{tab:galex_photometry}.

\begin{table*}[ht]
\centering
\caption{GALEX photometry results for TYC 4144-329-1 and TYC 4144-329-2.\label{tab:galex_photometry}}
\begin{tabular}{cccc}
\hline\hline
Source & NUV (2004) & NUV (2005) & FUV (2004)\\
       & mag & mag & mag \\
\hline
TYC 4144-329-1 & $16.39\pm 0.05$ & $16.40\pm 0.01$ & $<25.7$ \\
TYC 4144-329-2 & $15.31\pm 0.03$ & $15.80\pm 0.01$ & $19.1\pm 0.2$ \\
\hline
\end{tabular}
\end{table*}

\subsection{KWS lightcurves}
\label{sect:kwslightcurves}

The Kamogata/Kiso/Kyoto Wide-field Survey \citep[KWS,][]{oai:jaxa.repo.nii.ac.jp:00001963} provides $V$ and $I_c$ band data from about 2013 on. The KWS pipeline uses aperture diameters of 6 pixels (almost 2\arcmin) for aperture photometry in the $V$ band and 9 pixels (almost 3\arcmin) for the $I$ band. Thus, both members of the binary are safely inside the aperture at all times. Gaia and 2MASS images show that TYC 4144-329-1 and TYC 4144-329-2 are by far the brightest stars in that region, so any observed variability comes from either star or a combination of both.

\begin{figure*}
\plotone{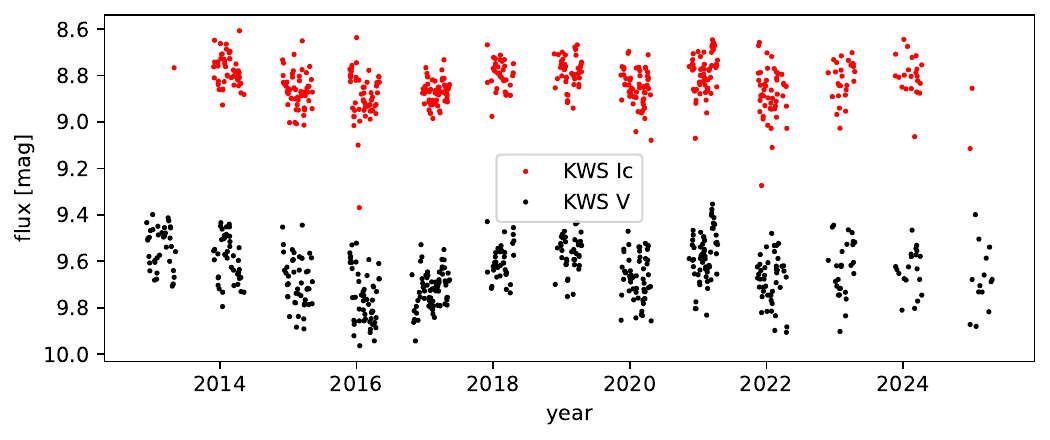}
\caption{Optical lightcurves of the unresolved binary TYC~4144-329-1 and TYC~4144-329-2 from KWS.
Error bars are omitted for clarity. 
An interactive version of this figure is available in the published journal article that combines the lightcurves in this figure with the data from figure~\ref{fig:lc_tess}.
\label{fig:lc_kws}}
\end{figure*}

We remove data points with uncertainties $>0.1$~mag and average multiple observations taken in the same night. Observations in the $V$ and $I_c$ bands are typically taken within 5~min, with a maximal time difference of 42~min. We assume that the target does not significantly change during this period and calculate the $V-I_c$ color for those nights where data in both bands is available. The light curve is shown in figure~\ref{fig:lc_kws}.

\subsection{TESS lightcurves}
\label{sect:tesslightcurves}

\begin{figure}
\plotone{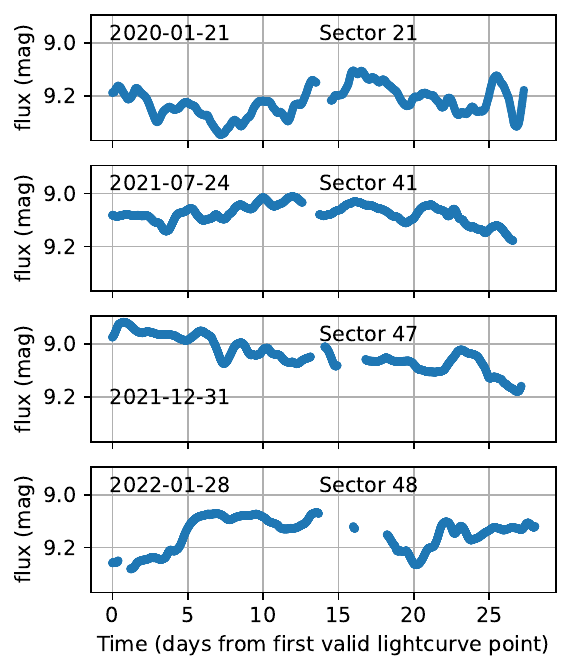}
\caption{Optical lightcurves of the unresolved binary TYC~4144-329-1 and TYC~4144-329-2 from TESS.
The date of the first valid lightcurve point for each sector is given in the labels.
\label{fig:lc_tess}}
\end{figure}
The Transiting Exoplanet Survey Satellite \citep[TESS,][]{2015JATIS...1a4003R} observed TYC 4144-329-1 and TYC 4144-329-2 in sectors 21, 41, 47, and 48. In sectors 74 and 75 the source is so close to the edge of the detector (or just beyond the edge) that no aperture photometry can be derived. We use the tessilator software \citep{2024MNRAS.533.2162B} to retrieve the TESS data \dataset[10.17909/0cp4-2j79]{http://dx.doi.org/10.17909/0cp4-2j79} and perform aperture photometry. The (unresolved) binary is by far the brightest source in the vicinity and contamination is not important. The tessilator uses a circular aperture and a background annulus and discards datapoints in the beginning and end of each sector where the background is not stable.

Figure~\ref{fig:lc_tess} shows the lightcurves. While the general flux level is similar in all sectors, consistent with the moderate long-term evolution seen in figure~\ref{fig:lc_kws}, there is also variability with a time scale of a few days.

\subsection{Other lightcurves}
\label{sect:otherlightcurves}

We investigated other archives for long-term lightcurve data, but the combined binary is too bright and saturated for most of them. DASCH \citep[Digital Access to a Sky Century @ Harvard,][]{ 2012IAUS..285...29G} provides access to digitized photographic plates from the Harvard College Observatory. Those observations are taken with different telescopes, instruments, and exposure times leading to large systematic uncertainties. For TYC 4144-329-2, observations reach back to 1899 and are consistent with a constant magnitude, albeit with uncertainties around 0.5~mag. Multi-year variability as seen in the KWS lightcurves (section~\ref{sect:kwslightcurves}) could be present but would not be discernible within the noise.

\subsection{Optical spectroscopy}
\label{sect:opticalspectroscopy}

Table~\ref{tab:observation_log} lists the optical spectra taken for TYC 4144-329-2. The Keck/HIRES \citep{1994SPIE.2198..362V} spectrum is described in \citet{2009ApJ...696.1964M}. 
TYC 4144-329-2 was also observed at Lick Observatory with the Shane 3\,m
telescope. Light was fed into the coud\'{e} focus which houses the
Hamilton echelle spectrograph \citep{1987PASP...99.1214V}. 
Data reduction for the Hamilton echelle with IRAF tasks is outlined in detail
in Lick Technical Report No.\ 74\footnote{\url{http://astronomy.nmsu.edu/cwc/Software/irafman/manual.html}}.
Briefly, data are bias subtracted, flat-fielded, extracted, and finally wavelength
calibrated with either ThAr or TiAr arclamp spectra (for the latter see \citealt{2013AJ....146...97P}).
\citet{2009ApJ...696.1964M} found a radial velocity of $-27.5\pm1.4$~km~s$^{-1}$ and we apply this velocity to all spectra.

The Hamilton spectrum in 2014 was taken with an iodine cell, all the other spectra without it. Before 2013, artifacts on the Hamilton detector like hot or cold pixels and columnar defects are present in many of the spectral orders. In the analysis, we concentrate on a few specific lines; those were checked for such artifacts and problematic regions are masked and interpolated.
The detector was upgraded between 2011 and 2013, leading a different format of the spectra with fewer defects. Unfortunately, the blaze function varies between datasets and it is not straightforward to properly correct it in each epoch. 
To have a homogeneous treatment of each epoch for robust variability assessment, we fit a simple analytical model of a Gaussian plus a constant to the continuum. 
The fit ignores lines or defects by sigma-clipping wavelength regions that deviate significantly from the fit, leaving us a with list of continuum points to determine the shape of the continuum. This results in a continuum normalization with visible residual patterns on scales $>100$~\AA{} at the 5\% level but where the continuum is flat on smaller scales.

\begin{table*}
\centering
\caption{Observation log for optical spectra.\label{tab:observation_log}}
\begin{tabular}{ccccccc}
\hline \hline
Telescope & Instrument &   UT obs mid-point   &  HJD            & resolution & exposure (s) & SNR (6500A) \\ 
\hline
Keck~I   &    HIRES   &  2007-05-05T09:27:07 &  2454225.89096  & 40,000    &  500     & 145 \\ 
Shane    &  Hamilton  &  2010-04-24T05:19:13 &  2455310.72206  & 39,000    & 2700     &  41 \\ 
Shane    &  Hamilton  &  2010-04-24T09:28:52 &  2455310.89541  & 39,000    & 2700     &  47 \\ 
Shane    &  Hamilton  &  2011-01-25T08:52:16 &  2455586.87351  & 39,000    & 2700     &  93 \\ 
Shane    &  Hamilton  &  2013-03-25T07:02:13 &  2456376.79545  & 39,000    & 2700     &  86 \\ 
Shane    &  Hamilton  &  2014-02-23T11:45:25 &  2456711.99341  & 62,000    & 1800     &  62 \\ 
Shane    &  Hamilton  &  2015-03-10T10:59:12 &  2457091.96077  & 62,000    & 1800     &  29 \\ 
Shane    &  Hamilton  &  2016-12-29T13:15:38 &  2457752.05585  & 62,000   & 1800     &  76 \\ 
\hline
\end{tabular}
\end{table*}
\begin{figure}
\plotone{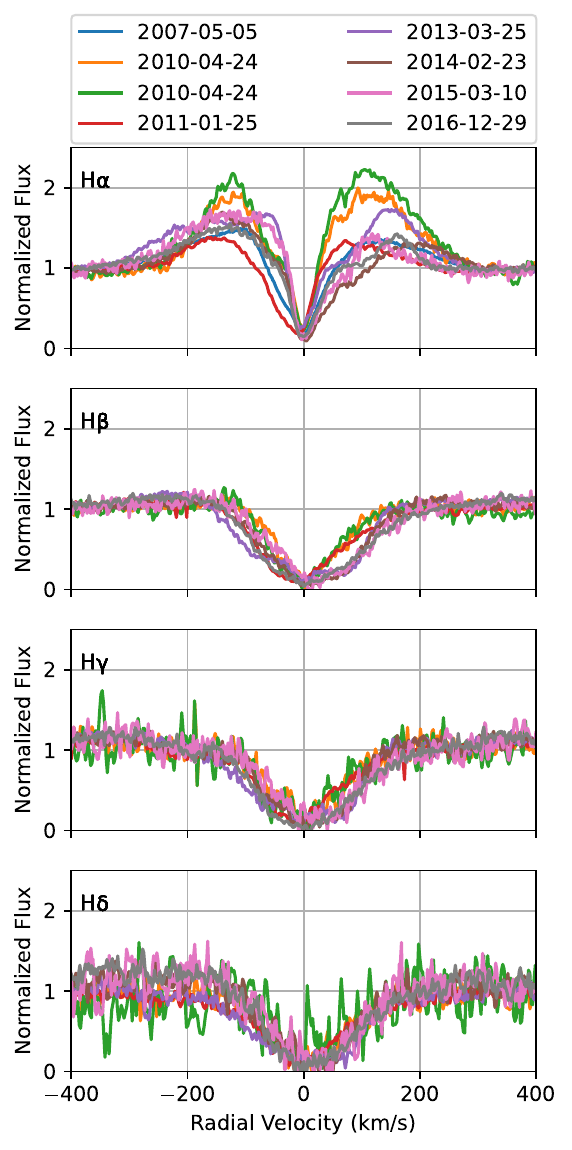}
\caption{Balmer line profiles of TYC 4144-329-2. 
There is considerable variability in both the flux and the line profile in H$\alpha$ but the other Balmer lines are more stable.
The spectra are continuum normalized.
\label{fig:spectra}}
\end{figure}
\section{Results}
\label{sect:results}
We first look at photometric and spectroscopic variability in the context of stellar activity (section~\ref{sect:opticalvariability} and section~\ref{sect:opticalspectra}).
Then, we determine the stellar activity in X-rays (section~\ref{sect:xrayfluxes}) and the UV (section~\ref{sect:uvfluxes}). We then try to predict the stellar structure (section~\ref{sect:stellarstructuremodels}) to explain the observed activity with a stellar dynamo (section~\ref{sect:chromosphericandcoronalactivity}).

\subsection{Optical variability}
\label{sect:opticalvariability}
Substantial variability is observed in optical lightcurves for the TYC 4144-329 system (Figures \ref{fig:lc_kws} and \ref{fig:lc_tess}).
Neither KWS nor TESS resolve TYC~4144-329-1 and TYC~4144-329-2, so we have to be careful to assign optical variability to one of the two stars. However, there are good reasons to expect that TYC~4144-329-2 is the variable star since we know it has a massive disk and indications of activity in the stellar spectrum as well as a larger value of $v\sin i$. Our GALEX data shows variability between observing epochs only in TYC~4144-329-2, and \citet{2009ApJ...696.1964M} also found $J$ band variability only in TYC~4144-329-2.

\subsubsection{Period}
\label{sect:period}
\begin{figure}
\plotone{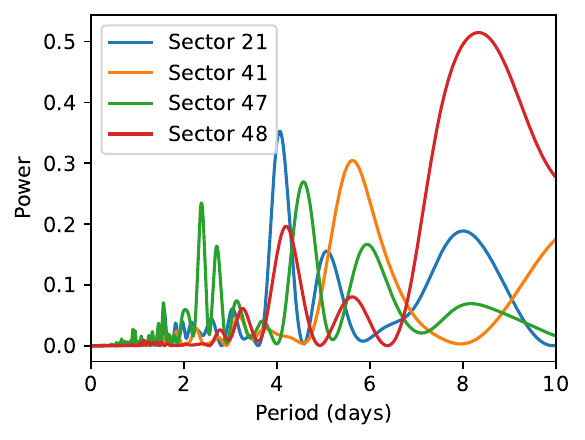}
\caption{Periodogram of the unresolved binary TYC~4144-329-1 and TYC~4144-329-2 from TESS.
\label{fig:period_tess}}
\end{figure}

Figure~\ref{fig:period_tess} shows the periodograms of the TESS light curves. While there is no single period that is significant in all sectors, there is a feature around 4-5 days that is noteworthy. It is the highest peak in sector 21 and 47; in sector 48 it is the second highest peak, with the highest peak seen at twice that period. Similarly, sector 47 shows a narrower double-peak at half this period. A tentative explanation could be a rotational period at either 4 or 8 days where we see an alias at twice the period. Sector 47 could be explained by a spot pattern of two spots that are separated by 180 degrees, while sector 21 happened to be observed at a period with irregular spot coverage. \citet{2009ApJ...696.1964M} saw signatures of accretion such as H$\alpha$ emission and a strong \ion{He}{1}~5877~\AA{} line in the spectrum of TYC 4144-329-2, so the surface features could either be related to magnetic activity or to accretion.

\citet{2009ApJ...696.1964M} estimate $v\sin i = 31$~km~s$^{-1}$ for TYC 4144-329-2, while TYC 4144-329-1 shows only $v\sin i = 3$~km~s$^{-1}$. From this, we can set an upper limit on the rotational period, using the surface gravity $\log g$ and the stellar mass $M_*$; with these numbers we find $P_\mathrm{rot}< 84$~d for TYC~4144-329-1 and $P_\mathrm{rot} < 4.4$~d for TYC~4144-329-2. The latter is consistent with a 4~day period in the light curves if the star is seen nearly edge-on.

We do not find any significant period in the KWS light curves but there are only 20-60 observations a band per observing season spread out over 8-9 months. If spots are present but their locations change over time, any periodic signal would be washed out over this range.

\subsubsection{Color variability}
\label{sect:colorvariability}

\begin{figure}
\plotone{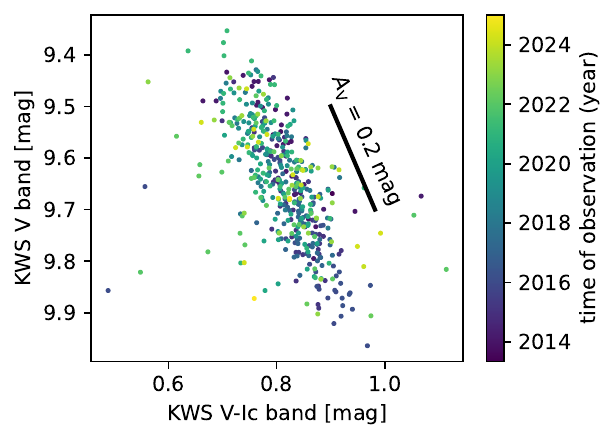}
\caption{Long-term color-magnitude diagram of the unresolved KWS lightcurve.
The observed distribution is consistent with reddening by dust
but could also be explained by spots on one of the stars.
\label{fig:CMD}}
\end{figure}

Figure~\ref{fig:CMD} shows the KWS color-magnitude diagram (CMD) of the combined photometry of the two stars. While there is significant scatter beyond the observational uncertainties, the data in general follows a reddening trend consistent with an ISM dust extinction law (using the tables in \citet{1998ApJ...500..525S} based on \citet{1989ApJ...345..245C}), which could point to clouds of dust particles from the disk passing through our line-of sight to TYC~4144-329-2. That is consistent with the analysis in \citet{2009ApJ...696.1964M}, who favor an ISM-like or slightly evolved dust extinction law and find about $A_\mathrm{V}=1.0$~mag for TYC~4144-329-2. The dip in brightness between 2015 and 2018 (figure~\ref{fig:lc_kws}) would then be a fraction of the total reddening. In this scenario, we probably see the disk close to edge-on and the ongoing accretion of material from the inner disk causes a change the density or geometric size of the disk, which lifts additional material into the line-of-sight.

Alternatively, cool spots on the stellar surface caused by magnetic activity would also make the star appear redder when it is fainter. The dip from about 2015-2018 could correspond to a period of increased magnetic activity on the star similar to the activity cycle on the Sun in this scenario. With long-term photometry in just two bands we cannot distinguish these scenarios though the latter one seems less likely given the lack of a single, deep convection zone in the MESA models (section~\ref{sect:stellarstructuremodels}).

\subsection{Optical spectra}
\label{sect:opticalspectra}

Figure \ref{fig:spectra} shows the H$\alpha$-H$\delta$ line profiles of TYC 4144-329-2 varying over time. In H$\alpha$, we see red and blue-shifted emission with a central absorption component. Most of the spectra can be fitted phenomenologically with a broad Gaussian emission component with peaks between -20 and +20~km~s$^{-1}$ and a centered, narrow absorption component. However, there are significant deviations from this simple model most significantly on 2013-03-25 when the spectrum shows a second, weaker absorption component that is slightly redshifted. On the blue side the line has three distinct peaks between $-200$ and $-50$~km~s$^{-1}$. The two spectra from 2010-04-24 show a second, weaker absorption component that is blue-shifted. The two spectra are taken about four hours apart and in that time the flux between $+50$ and $+200$~km~s$^{-1}$ increases by about 20\%.

The higher-order Balmer lines are simpler, less variable, and also increasingly noisy. There is a hint of the weak, blue-shifted emission component in H$\beta$ in some spectra, but the line is dominated by a single absorption component. In the spectra from 2010 and 2011 (orange, green, and red) the line is almost symmetric, while the other spectra have a significantly deeper profile in the +0 to +100~km~s$^{-1}$ range. It remains unclear if this is due to additional emission filling in this region in 2010 and 2011 or additional absorption in the other spectra. A comparison with H$\alpha$, where the 2010 spectra show emission above the continuum level in this region, makes it plausible that this might be a real emission component. Either way, we can conclude that most of the variability is only present in H$\alpha$. 
The equivalent width of H$\alpha$ varies between -0.4 and -5~\AA{}, so overall the line is always in emission.

We do not see variability in other lines that are commonly associated with stellar activity on cool stars, such as the Ca~II H\&K lines or the Ca~II infrared triplet, nor do these lines have indications of narrow emission cores.

\subsection{X-ray fluxes}
\label{sect:xrayfluxes}

Converting the upper limit on the flux of TYC~4144-329-1 and the two counts of TYC~4144-329-2 to an energy flux requires a spectral model, since Chandra/ACIS has different sensitivity at different energies. 
We calculate responses and effective area and test different spectral models.
We estimate $A_V=1.1$~mag for TYC~4144-329-2 (table~\ref{tab:stellar_parameters}), assuming absorption by interstellar grains which corresponds to a hydrogen column density of $N_H=2.0\times 10^{21}$~cm$^{-2}$ \citep{2003A&A...408..581V}. 

With that absorbing column density, we find unabsorbed fluxes in the 0.5-7.0~keV band between $7\times 10^{-15}\;\mathrm{erg}\;\mathrm{cm}^{-2}\;\mathrm{s}^{-1}$ and $1\times10^{-15}\;\mathrm{erg}\;\mathrm{cm}^{-2}\;\mathrm{s}^{-1}$ for coronal temperatures between 0.4 and 1.5~keV, which corresponds to X-ray luminosities of $L_X=1\times 10^{29}\;\mathrm{erg}\;\mathrm{s}^{-1}$ and $2\times 10^{28}\;\mathrm{erg}\;\mathrm{s}^{-1}$, respectively. The $A_\mathrm{V}$ value for TYC~4144-329-1 corresponds to $N_H=7\times10^{17}\;\mathrm{cm}^{-2}$ (see table~\ref{tab:stellar_parameters}), which leads to upper limits on the unabsorbed flux of $L_X=8\times 10^{28}\;\mathrm{erg}\;\mathrm{s}^{-1}$ and $2\times 10^{28}\;\mathrm{erg}\;\mathrm{s}^{-1}$, respectively for the two temperatures.

\subsection{UV fluxes}
\label{sect:uvfluxes}

Table~\ref{tab:galex_photometry} shows that TYC~4144-329-1 has a consistent NUV flux between 2004 and 2005 but TYC~4144-329-2 dims by half a magnitude. In order to derive energy fluxes in the UV, we need to deredden the observed fluxes; for the GALEX bands we use the empirical coefficients from \citet{2013MNRAS.430.2188Y} which are derived from Galactic stars with spectral types similar to TYC~4144-329-1 and TYC~4144-329-2. We use the $A_\mathrm{V}$ values from table~\ref{tab:stellar_parameters} but caution that time variable extinction can change derived fluxes significantly.

For the 2004 GALEX observations where both FUV and NUV bands were observed, we find an upper limit in the FUV of $<4\times 10^{-16}\;\mathrm{erg}\;\mathrm{cm}^{-2}\;\mathrm{s}^{-1}$ for TYC~4144-329-1 and an NUV flux of $6\times10^{-12}\;\mathrm{erg}\;\mathrm{cm}^{-2}\;\mathrm{s}^{-1}$, which corresponds to UV luminosities of $L_\mathrm{FUV}<7\times 10^{27}\;\mathrm{erg}\;\mathrm{s}^{-1}$ and $L_\mathrm{NUV}=1\times 10^{32}\;\mathrm{erg}\;\mathrm{s}^{-1}$, respectively. For TYC~4144-329-2, the dereddened NUV flux estimated this way would be larger than the bolometric flux given in table~\ref{tab:stellar_parameters}, indicating that the extinction was probably lower at the time of the GALEX observations, the UV photons are scattered in the disk, or some fraction of the UV flux is not subject to the same reddening as the star itself. This is possible in certain geometric configurations. If the star is observed close to edge-on, the disk would obscure the stellar light, but X-ray and UV emission might originate in the corona at a larger height and be seen above the disk or might scatter off the disk rim to reach us indirectly.

\subsection{Stellar structure models}
\label{sect:stellarstructuremodels}

In the Sun and other late-type stars, the outer convective zone powers dynamo action which ultimately leads to the formation of a stellar magnetic field which causes stellar spots and heats the corona. One way to describe the convection is through the Rossby number $Ro=\frac{P_\mathrm{rot}}{2 \pi}\frac{v_\mathrm{conv}}{H}$ where $P_\mathrm{rot}$ is the rotation period, $v_\mathrm{conv}$ is the convective velocity, and $H$ is the radial extent of the convection zone. The Rossby number describes the ratio between rotation and convective turn-over time. 

$v_\mathrm{conv}$ and $H$ are not directly observable, so they have to be derived from models. To that end, we run MESA \citep[Modules for Experiments in Stellar Astrophysics,][]{2011ApJS..192....3P,2013ApJS..208....4P,2015ApJS..220...15P,2018ApJS..234...34P,2019ApJS..243...10P,2023ApJS..265...15J} models with software version r24.08.1. The models do not include the effects of a merger or accretion, nor the slow stellar rotation. 
Stellar models are evolved from the main sequence with a specified mass until they reach a given radius, at which point the radial extent of the outer convective zone and the maximum convective velocity in the zone are used to compute the Rossby number. This differs from the approach by \citet{2020Natur.587..387H} who injected a burst of energy into the envelope at the point of the merger and then continued to evolve the star from that point. In their simulations the state of the atmosphere settles into a new equilibrium after a few hundred years.

For comparison, we calculate models for the Sun and a few other stars. Helioseismic measurements put the Rossby number in the Sun at about 5 \citep{2016ApJ...824....4G}. The Sun has a similar $\log{L_\mathrm{X}/L_\mathrm{bol}}$ as TYC~4144-329-2 (we use fluxes from \citet{2000ApJ...528..537P} for the solar minimum and maximum). We also run models for $\psi^3$~Psc, a F9IIIa giant \citep[see][for stellar parameters]{Gondoin_2005}, again with a similar $\log{L_\mathrm{X}/L_\mathrm{bol}}$ as TYC~4144-329-2, and FK~Com, thought to be the end stage of a stellar merger, with a much higher $\log{L_\mathrm{X}/L_\mathrm{bol}}$ and a more contracted atmosphere compared to TYC~4144-329-2. FK~Com's rotation period is only 2.4~days \citep{2013A&A...553A..40H} with $M_*=1.1\;M_\odot$ \citep{2019A&A...628A..94A} and $R=6.99\;R_\odot$ \citep{2018A&A...616A...1G}.

\begin{figure}
\plotone{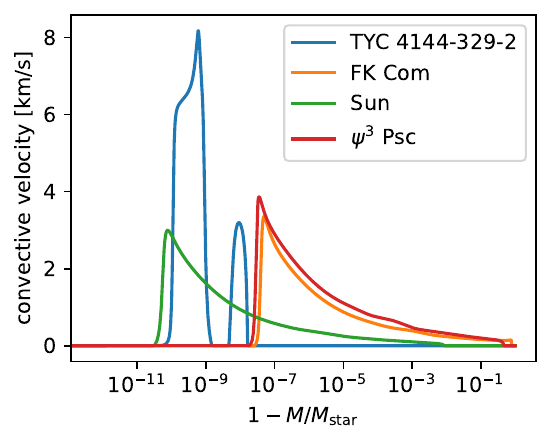}
\caption{Convective velocity in the outermost layers of the stellar atmosphere from MESA models.
The x-axis shows the mass, from the outside towards the center of the star.
TYC~4144-329-2 stands out with two distinct, yet very thin convection zones.
\label{fig:mesa}}
\end{figure}

Figure~\ref{fig:mesa} shows the locations of the convection zone in those models. The Sun, FK~Com, and $\psi^3$~Psc all have a single convection zone with the highest convection speed close to the surface and falling off gradually towards the interior. In contrast, the model predicts two distinct, but thin convection zones for TYC~4144-329-2. Given the small dimension and the simplifications in the model setup the exact numbers should be interpreted carefully, but it is obvious that the structure of the atmosphere is fundamentally different. Using the maximal convective speed and just the extent of the outer convection zone for TYC~4144-329-2, we find a Rossby number of 96, while the numbers for the other stars are close to previously published values with a 6.9 for the Sun \citep[about 5,][]{2016ApJ...824....4G} and 0.1 for $\psi^3$~Psc \citep[0.17,][]{Gondoin_2005}.

\subsection{Chromospheric and coronal activity}
\label{sect:chromosphericandcoronalactivity}

\begin{figure*}
\plotone{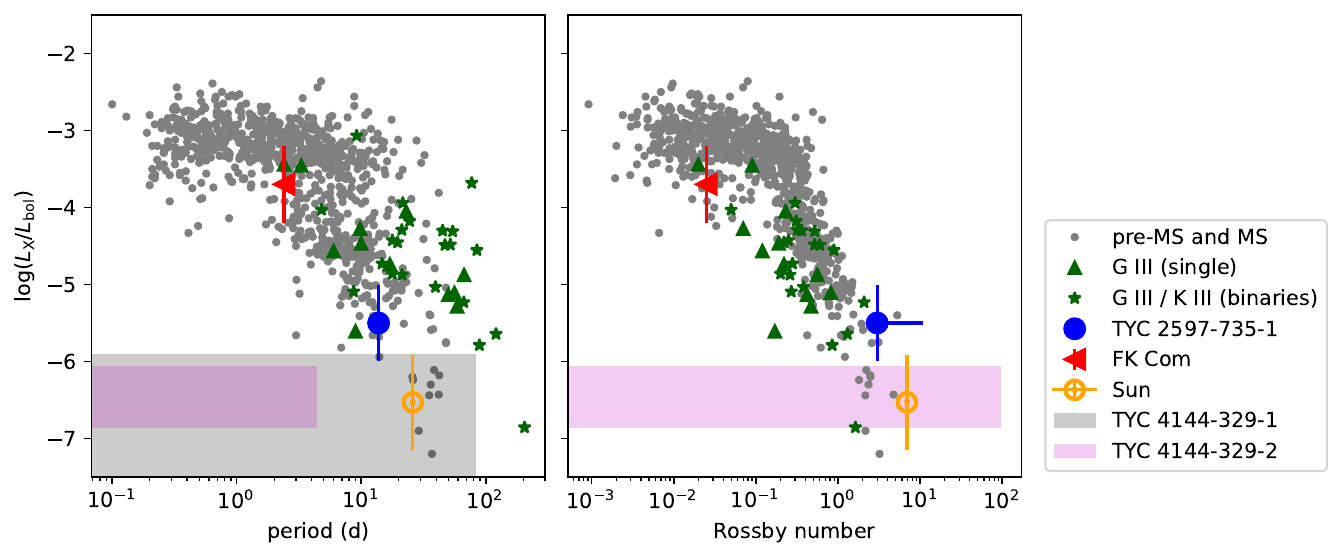}
\caption{\emph{left}: For coronal sources, the ratio of X-ray luminosity ($L_\mathrm{X}$) and bolometric luminosity $L_\mathrm{bol}$ is correlated with the rotational period. 
The figure compares the limits on TYC~4144-329-1 and TYC~4144-329-2 with other stellar samples; 
see ~\ref{sect:chromosphericandcoronalactivity} for a description of the samples.
Error bars are omitted for samples with more than three objects for clarity.
Since the inclination of TYC~4144-329-1 and TYC~4144-329-2 is unknown, there is only an upper limit on the rotational period. 
TYC~4144-329-1 is undetected in X-rays; the area marked in the plot assumes the least restrictive coronal temperature (0.3~keV) discussed in the text.
For hotter coronae, the limit on the X-ray luminosity is lower.
\emph{right}: Same as left panel but for the Rossby number.
\label{fig:lxlbol}}
\end{figure*}

We did not detect X-rays from TYC~4144-329-1 and set a limit on the ratio of X-ray and bolometric flux of $\log{L_\mathrm{X}/L_\mathrm{bol}}<-5.8$ assuming a very cool corona and a reasonable column density. For hotter coronae, that limit is even lower. At the age we determine for the system, such low X-ray activity is expected \citep{2017MNRAS.471.1012B}.  Figure~\ref{fig:lxlbol} shows the X-ray activity of TYC~4144-329-1 in the context of other stars with known rotation periods and X-ray luminosities. The gray dots in the figure are dwarf stars from the sample of \citet{Wright_2011}, which includes both main-sequence and pre-main-sequence stars. Fast rotating stars (often young) saturate around $\log{L_\mathrm{X}/L_\mathrm{bol}}\approx -3$, and as stars age and lose angular momentum, their X-ray luminosity decreases. 


The gray area in the figure marks the region compatible with the limits for TYC~4144-329-1. If we see the star close to edge-on, i.e.\ the rotation period is long, then the X-ray limit is similar to the detections of the slowest rotating stars in the sample of \citet{Wright_2011}. 

Figure~\ref{fig:lxlbol} also shows a sample of single giant G-type stars \citep{Gondoin_2005} and G-type giants that are part of binary systems \citep{Gondoin_2007}. They follow a very similar trend as main-squence dwarf stars, except that none of them are rotating fast enough to be in the saturated regime. 

The left panel compares $L_\mathrm{X}/L_\mathrm{bol}$ to the rotation period and the right panel compares $L_\mathrm{X}/L_\mathrm{bol}$ to the Rossby number. The right panel contains fewer objects, because the Rossby number is unknown for some of them. It is also worth noting the Rossby number in the samples from \citet{Wright_2011} is based on an empirically fitted relation while \citet{Gondoin_2005,Gondoin_2007} uses a fit to a grid of models from \citet{1998MNRAS.296..150G}. The Rossby number for the individual objects we discuss in this work are based on the MESA models specifically tailored to each object (section~\ref{sect:stellarstructuremodels}). Yet, all points follow the same general trend.

Our X-ray observations test if TYC~4144-329-2 is intermediate in evolution between stars that likely just accreted a companion and FK~Com-type stars. 
TYC~2597-735-1 shows signs of a very recent accretion event. It is surrounded by an UV-bright nebula, presumably formed by an outflow caused by the recent break-up and accretion of a companion \citep{2020Natur.587..387H}. X-ray emission from the central source was observed by \citet{2022AJ....163..173G} who also found weak emission from points in the nebula, possibly from shocks in the outflow, which point to relatively fast outflow speeds and thus a relatively recent merger event. Again, this leads us to believe that TYC~2597-735-1 is less evolved than TYC~4144-329-2 and thus should be X-ray fainter. On the other hand, FK~Com is thought be the end product of the binary merger process, a star that is rapidly rotating and shows many signs of activity, including hot, chromospheric emission lines and strong X-ray emission \citep{2016ApJS..223....5A}.

However, figure \ref{fig:lxlbol} shows that TYC~4144-329-2 is X-ray fainter than TYC~2597-735-1 and FK~Com although TYC~4144-329-2's rotation period must be shorter than the period of TYC~2597-735-1.
\citet{2022AJ....163..173G} discussed other possible mechanisms for the X-ray emission of TYC~2597-735-1, including shocks in the outflow, but they conclude that the observed limits on outflow speed and mass loss rate are restricting this scenario and a coronal origin is more likely. 
In TYC~4144-329-2, we do not see a jet or ring nebula that could contribute to the X-ray emission. At the same time, the energy of the two detected photons is well above 1~keV, which would require shock speeds at least twice as high as the free-fall speed \citep[e.g.][]{2007A&A...466.1111G} leaving a corona as the obvious source of the weak emission. The fact that TYC~4144-329-2 is X-ray fainter than the other suspected merger remnants thus can only be explained by having a different atmospheric structure not conducive to producing magnetic fields via a convective dynamo. In fact, TYC~4144-329-2 is less X-ray active than MS stars or giants with a similar rotation period. Given how well X-ray activity is generally correlated with rotation, this indicates a scenario where the accretion of a companion actually disrupts the magnetic dynamo in some way or the star is in a peculiar place in its evolution. The MESA simulations in section~\ref{sect:stellarstructuremodels} point to this latter scenario with two separate and very narrow convection zones instead of a single, extended zone.

In the UV, we can look at the GALEX FUV-NUV color in comparison to the effective temperature. \citet{2014AJ....147..159S} analyzed stars with SDSS and GALEX observations in the Kepler field and found a population of UV excess stars which they associate with active chromospheres \citep[figure 3 in ][]{2014AJ....147..159S}. TYC~4144-329-1 is inactive by this measure, and so is TYC~4144-329-2 with its FUV-NUV color between 3.8 and 6.1, dereddened with $A_V=0$~mag and $A_V=1$~mag, respectively. 

\section{Discussion}
\label{sect:discussion}

With the new Gaia distance, we derived the stellar parameters for TYC~4144-329-1 and TYC~4144-329-2. TYC~4144-329-1 is an evolved star with an age of $2.4\pm0.6$~Gyr that we can use to anchor the age of the system. TYC~4144-329-2 has several unusual features, most notably a massive dusty disk and signs of accretion in the stellar spectrum. We discuss our observations in the context of magnetic activity (section~\ref{sect:magneticactivity}) and accretion (section~\ref{sect:accretion}) to develop a consistent scenario for TYC~4144-329-2 (section~\ref{sect:scenario}). Based on that, we discuss possible future work (section~\ref{sect:outlook}).

\subsection{Magnetic activity}
\label{sect:magneticactivity}
While the X-ray flux from TYC~4144-329-2 is low, the star is clearly detected and the energy of the two photons is too high to be formed in outflows or accretion streams (section~\ref{sect:xrayfluxes}), pointing to coronal emission at a very low activity level. The corona is much fainter than one would expect from the rotation period but compatible with other sources once we take the small extent of the convection zone into account and look at the $L_X/L_\mathrm{bol}$ to Rossby number relation (figure~\ref{fig:lxlbol}). The $FUV-NUV$ color is also consistent with a very inactive star, and neither the Ca~{\sc ii} H\&K lines nor the Ca~{\sc ii} infrared triplet show any indications of magnetic activity, but a few star spots might be present and cause or at least contribute to the observed variability in the optical and UV. So, what can we say about the magnetic activity of TYC~4144-329-2 beyond the fact that it is weak?

\citet{1997A&A...318..215S} collected a complete sample of A, F, and G dwarfs in the solar neighborhood observed with \emph{ROSAT}. He finds an apparent minimum magnetic activity level for all spectral types where the X-ray surface flux is comparable to the flux observed in a solar coronal hole around $10^4$~erg~s$^{-1}$~cm$^{-2}$. The upper limits on the X-ray surface flux for TYC~4144-329-1 is $F_X<3-16\times10^4$~erg~s$^{-1}$~cm$^{-2}$, depending on the assumed coronal temperature and converted into the ROSAT/PSPC band (0.1-2.4 keV) for comparison with \citet{1997A&A...318..215S}; for TYC~4144-329-2 we find $F_X=1.4-18\times10^5$~erg~s$^{-1}$~cm$^{-2}$. The flux from TYC~4144-329-1 is just consistent with that idea of a minimum level, while TYC~4144-329-2 is a factor of a few brighter. So we can imagine that the entire surface of TYC~4144-329-1 looks like a coronal hole on the Sun, while TYC~4144-329-2 has a few active regions.

From the $L_X/L_\mathrm{bol}$ we can also estimate the magnetic field strength. \citet{2014MNRAS.441.2361V} looked at about 100 magnetic surface maps from Zeeman Doppler imaging and find relations between age, rotation, X-ray luminosity, and magnetic field strength. Relations for age and rotation are not useful for TYC~4144-329-2 because the suspected recent stellar merger and accretion from the disk will have spun up the star compared to an unperturbed evolution but they also observe a lower magnetic field strength with increasing Rossby number and lower $L_X/L_\mathrm{bol}$. There is significant scatter in the relation and the convection zone in TYC~4144-329-2 looks very different from convective main-sequence stars, so it is not clear that those relations hold for TYC~4144-329-2, but if they do, it suggests a large scale magnetic field $<1$~G.

\subsection{Accretion}
\label{sect:accretion}

\citet{2009ApJ...696.1964M} already pointed to the H$\alpha$ emission and the strong \ion{He}{1}~5877~\AA{} line as signs of ongoing accretion from a circumstellar disk observed in the IR. The variability in the H$\alpha$ line profile (section~\ref{sect:opticalspectra}) supports this view as the spectra are best understood as a combination of emission and absorption components that will occur when accreting matter passes through the line-of-sight.

We also see changes in absorption on much shorter time scales within every optical observing season (see the CMD in figure~\ref{fig:CMD}) of $\Delta A_\mathrm{V}=0.2-0.4$~mag. The same figure also shows considerable scatter around the absorption trend beyond the observational uncertainties. While some of that scatter could be due to changes in the dust composition of the inner disk, spots that rotate in and out of view can change the observed color as well. The TESS lightcurves indicate a period around 4 or 8 days. If the scatter is caused by spots, then this period is the stellar rotation period; if it is caused by the accretion streams or inner disk features it points to an inner disk radius of 5 or 7 stellar radii. In young stars with their kG magnetic fields the inner disk is magnetically connected to the star and thus the inner disk radius tends to be close to the co-rotation radius \citep[e.g.][]{1994ApJ...429..781S}.  
However, the low X-ray activity of TYC~4144-329-2 indicates the absence of strong large-scale magnetic field so we cannot assume that scaling laws between H$\alpha$ line emission and accretion rate derived from magnetically funneled accretion onto young stars
hold for TYC~4144-329-2 or that the optical period represents the stellar rotation period. 

Instead, accretion could proceed through a boundary layer where mass slowly spirals in just below the Keplerian velocity and then decelerates rapidly in a thin boundary layer as it comes into contact with the stellar surface. Boundary layers have been studied in the past in the context of young stars \citep[e.g.][]{1988ApJ...330..350B} before the magnetically funneled accretion model took hold, but they are common in other astrophysical objects with weak magnetic fields like planets, very early proto-stars, and some white dwarfs. Boundary layers do not reach the same temperature as accretion shocks, because the gravitational energy is not released in a single shock surface, but over the range of the boundary layer. Mass and angular momentum are carried through waves, shocks, and mass in- and outflows both into the star and out into the disk \citep[][]{2022MNRAS.512.2945C,2025ApJ...985...16T}. The Keplerian velocity at the stellar surface of TYC~4144-329-2 is about 350~km~s$^{-1}$, similar to the width of the wings of the H$\alpha$ line. When the disk reaches close to the stellar surface, it blocks the view to the far side, but if the star is seen close to edge on, we would observe emission from the boundary layer both moving towards and away from us. Accretion in a shock or boundary layer would also cause bright UV emission but if the same dust column density applies to the UV as to the optical, deredding the observed UV emission leads to a flux above the bolometric luminosity (section~\ref{sect:uvfluxes}) which is unphysical. Instead, the UV emission could be produced elsewhere (e.g.\ in stellar activity) or be scattered off the disk such that absorption for the UV light path is less than along the line-of-sight to the star itself.

We know that the disk is not flat given its high IR luminosity compared to the stellar luminosity \citep{2009ApJ...696.1964M}, so there can still be clumps and other features that rotate into and out of the line-of-sight and change the $A_V$ over time.
The accretion must be weak since the equivalent width of H$\alpha$ never exceeds -5~\AA{}, and is usually much weaker. So, while the outer disk seems massive in the IR, there might be an inner gap that limits the mass flow. Such a gap could be opened up by another body in the system (\citet{2003ASPC..293...76W} suggests that the core of the companion remains after stripping most if its mass) or by the stellar irradiation. A future study of the inner disk properties based on IR spectroscopy can characterize the inner disk and reveal why the accretion rate is so low, given the strong IR excess of the disk.

\subsection{A consistent scenario for TYC~4144-329-2}
\label{sect:aconsistentscenariofortyc41443292}

\label{sect:scenario}
The X-ray and UV data from TYC~4144-329-2 indicate a star with a low magnetic activity level, but the variability and spectral tracers suggest ongoing accretion from the massive circumstellar disk. 
Together with the relatively high rotational velocity for an evolved star, we can paint a consistent picture where TYC~4144-329-2 is observed at an intermediate inclination, and the line-of-sight passes through the upper layers of a flared disk, which reaches down to the stellar surface where it forms an accretion boundary layer. As the disk rotates and disk structure changes over time, the absorbing column density changes. Similar effects have been observed in young stars with primordial accretion disks \citep[e.g.][]{2015A&A...584A..51S,2018AJ....156...56G}. If increased absorption is indeed responsible for the lower optical flux between 2015 and 2018, then the thicker regions of the flared disk that rotated through the line-of-sight must be located at a radius beyond 6~au since only one such long-lived absorption event has been seen in the KWS data that span more than a decade. Alternatively, we could be witnessing a cycle in stellar activity. 

The UV emission is also time variable, but likely does not suffer the same extinction as the photosphere. That could simply be an effect of time variability since the UV data was observed about a decade before the KWS light curve started. It could also point to a geometry where the UV emission is scattered or originates in a corona above the stellar surface so that the line-of-sight for the UV emission has a lower average absorption than the optical light. 

Where does that leave TYC~4144-329-2 in comparison to other stellar merger candidates? As it accretes more mass and angular momentum from the disk, TYC~4144-329-2 might spin up further and possibly also contract towards a fast rotating, highly active state with a deeper convection zone like FK~Com. 
Compared to other stellar merger candidates with disks, in particular TYC~2597-735-1, no jet or wide angle outflows has clearly been detected around TYC~4144-329-2 which could be interpreted as a later stage of disk accretion. However, the FUV data is also consistent with slightly extended emission (section~\ref{sect:uvdata}). For TYC~2597-735-1 \citet{2020Natur.587..387H} present a grid of MESA models, also with two convection zones. \citet{2022AJ....163..173G} derive Rossby numbers for this grid and find that the Rossby number is initially very large, because the star is inflated from the accretion, but decreases as the star contracts post-merger to a range where we would expect more magnetic activity after a few thousand years. Thus, the merger event in TYC~4144-329-2 might actually be more recent than in TYC~2597-735-1 and jets and winds might form later, when the star begins to contract again, the convection zone deepens, and the magnetic fields becomes stronger.

This timeline contrasts with \object{V838 Mon} which brightened by 9 magnitudes in 2002 and has also been attributed to a stellar merger \citep{2006A&A...451..223T}. However, V838~Mon appears as a red giant today \citep{2025MNRAS.541.3331G} with ejections of dusty material \citep{2007A&A...467..269G} and possibly jets \citep{2024A&A...686A.260M}. For V838~Mon, \citet{2007MNRAS.375..909S} predict that the X-ray luminosity should peak about 6-8 years after and then decay over about a century after the merger event. Transient X-ray emission in the vicinity of the object has been detected \citep{2010ApJ...717..795A}, possibly from the ejecta. In contrast, TYC~4144-329-2 is hotter, but less luminous and shows a disk, but no known outflows. We also know from the DASCH light curve (section~\ref{sect:otherlightcurves}) that no bright outburst occurred in the last 125 years. We conclude that V838~Mon and similar objects like V1309~Sco are not just younger, but physically different from TYC~4144-329-2 and TYC~2597-735-1. We speculate that accretion of a companion might have been more gradual in TYC~4144-329-2 and TYC~2597-735-1 which allowed a disk to form, while V838~Mon swallowed the companion in a single, massive outburst. The difference might be due to the size and mass of the accreted companion and the mass and evolutionary stage of the primary at the time of the merger.

\subsection{Outlook}
\label{sect:outlook}
Our approximation of single star evolution in the MESA model (section~\ref{sect:stellarstructuremodels}) is clearly too simple, and \citet{2020Natur.587..387H} approach of injecting a single burst of energy also implies very strong assumptions on how the merging process between two stars actually works. \citet{2006A&A...451..223T} did an exploration of V838~Mon type mergers, based on a parametrized code from \citet{2002ApJ...568..939L} mostly aimed as understanding the outburst phase itself, which likely happened centuries or millenia ago for TYC~4144-329-2.
Simulations along the lines of what \citet{2024ApJ...962..168S} did for more massive stars could lead to significant progress in our understanding of TYC~4144-329-2 and similar systems such as TYC~2597-735-1 bringing us closer to quantifying the conditions under which non-degenerate binary stars merge, how they evolve after, and how that changes the populations of galaxies.

We also encourage further work on comparing the relative extent of optical variation or spots versus X-ray emission. We detect optical variability in TYC~4144-329-2 which might be due to spots, yet the $L_\mathrm{X}/L_\mathrm{bol}$ value is consistent with an inactive star despite the MESA simulations showing some convection in narrow and deep zones. Maybe a strong toroidal magnetic in the convections zone field still buckles, and loops rise buoyantly to the surface where it causes cools spots, but since the footpoints are embedded deeper in the star, there may be less field twisting or reconnection might occur deeper in the atmosphere. This scenario could also be explored in future simulations and be constrained observationally with more data from stars with convention zones with unusual properties such as TYC~4144-329-2.

\subsection{Reproducability}
\label{sect:reproducability}
Code and data to reproduce the results in this paper can be found at the following DOI: 10.5281/zenodo.22665084

\section{Summary}
\label{sect:summary}

We study the activity of TYC~4144-329-2 from X-rays to the optical. The star is part of a wide binary system with TYC~4144-329-1, which appears to be inactive, while TYC~4144-329-2 shows signs of weak, but persistent coronal activity resulting in a weak X-ray and FUV detection. \citet{2009ApJ...696.1964M} found a massive disk around TYC~4144-329-2 in the IR and identified this star as the remnant of a stellar merger. Other stellar merger candidates spin up as the result of the momentum accreted when a binary companion falls in, which leads to stronger activity. Our modeling indicates that TYC~4144-329-2 has only a very thin outer convection zone, so it is not able to generate strong magnetic fields. We speculate that TYC~4144-329-2 is in a very early stage of post-merger evolution, where the star is still inflated and has not yet contracted and did not yet have time to form a deep convection zone. 

We also find variability in H$\alpha$ line profiles and in the optical light curves. We suggest that an intermediate geometry where the line-of-sight passes through the upper layers of a flared disk can explain the observed photometric variability as time variable absorption, while a boundary layer at the stellar surface with a low and variable mass accretion rate causes the observed H$\alpha$ line profiles.

\begin{acknowledgments}
The scientific results reported in this article are based on observations made by the Chandra X-ray Observatory.
Support for this work was provided by the National Aeronautics and Space Administration through Chandra Award Number GO3-24005X issued by the Chandra X-ray Center, which is operated by the Smithsonian Astrophysical Observatory for and on behalf of the National Aeronautics Space Administration under contract NAS8-03060.
This publication makes use of data products from the Two Micron All Sky Survey, which is a joint project of the University of Massachusetts and the Infrared Processing and Analysis Center/California Institute of Technology, funded by the National Aeronautics and Space Administration and the National Science Foundation.
This work has made use of data from the European Space Agency (ESA) mission
{\it Gaia} (\url{https://www.cosmos.esa.int/gaia}), processed by the {\it Gaia}
Data Processing and Analysis Consortium (DPAC,
\url{https://www.cosmos.esa.int/web/gaia/dpac/consortium}). Funding for the DPAC
has been provided by national institutions, in particular the institutions
participating in the {\it Gaia} Multilateral Agreement.
C. M. is grateful for support in conducting optical spectroscopic observations through National Science Foundation award No. AST-1003318. 
Research at Lick Observatory is partially supported by a generous gift from Google.

\end{acknowledgments}

\begin{contribution}

HMG reduced and analyzed the X-ray and UV data, the optical light curves, and wrote most of the manuscript.
KS ran the MESA models, KS and EB contributed to the interpretation of the stellar structure.
CM observed and reduced the optical spectroscopy and context from earlier analysis.
All other authors contributed to the interpretation of this object and its astrophysical context.


\end{contribution}

\facilities{CXO (ACIS), GALEX, TESS, Shane (Hamilton), Keck:I (HIRES)}
\software{astropy \citep{2013A&A...558A..33A,2018AJ....156..123A,2022ApJ...935..167A},  
          CIAO \citep{2006SPIE.6270E..1VF,2026arXiv260514144F}, 
          Sherpa \citep{2024ApJS..274...43S},
          gPhoton \citep{2016ApJ...833..292M},
          tessilator \citep{2024MNRAS.533.2162B},
          Astrocut \citep{2019ascl.soft05007B},
          gPhoton \citep{2016ApJ...833..292M},
          insochrones \citep{2015ascl.soft03010M},
          MultiNest \citep{Buchner+2014},
          daschlab \citep{williams_2024_14574817},
          Github Copilot version 0.48.1,
          TESSCut \citep{2019ascl.soft05007B},
          }

\bibliography{bib}{}
\bibliographystyle{aasjournalv7}


\end{document}